\documentclass[twocolumn,letterpaper,aps,prd,longbibliography,superscriptaddress,showpacs,nofootinbib,floatfix]{revtex4-1}

\usepackage{xspace}
\usepackage{graphicx}

\newcommand{\pt} {\mbox{$p_T$}\xspace}

\newcommand{\acc} {\mbox{${\rm A}\epsilon_{\rm rec}$}\xspace}
\newcommand{\massmumu} {\mbox{$M_{\mu\mu}$}\xspace}

\newcommand{\dsigdydpt}{\frac{{\rm d}^2\sigma_\phi}{ {\rm d}p_T \; {\rm d}y }}

\newcommand{\s}{\ensuremath{\sqrt{s}}\xspace}

\newcommand{\pp}{\mbox{$p$+$p$}\xspace}
\newcommand{\pphi}{\mbox{$\phi$}\xspace}
\newcommand{\rhomega}{\mbox{($\omega+\rho$)}\xspace}

\newcommand{\mumu}{\ensuremath{\mu^+\mu^-}\xspace}

\newcommand{\GeV}{\ensuremath{{\rm ~GeV}}\xspace}

\newcommand{\GeVc}{\ensuremath{{\rm ~GeV}/c}\xspace}
\newcommand{\GeVcc}{\ensuremath{{\rm ~GeV}/c^2}\xspace}

\newcommand{\tab}[1]{Table~\ref{#1}}

\newcommand{\pythiasix}{{\sc pythia6}\xspace}
\newcommand{\pythiaeight}{{\sc pythia8}\xspace}
\newcommand{\pythia}{{\sc pythia}\xspace}
\newcommand{\phojet}{{\sc phojet}\xspace}
\newcommand{\epos}{{\sc epos3}\xspace}
\newcommand{\eposlhc}{{\sc epos-lhc}\xspace}
\newcommand{\ampt}{{\sc ampt}\xspace}
\newcommand{\urqmd}{{\sc urqmd}\xspace}
\newcommand{\geant}{{\sc geant3}\xspace}
\newcommand{\hijing}{{\sc hijing}\xspace}

\begin{document}

\title{Measurement of $\phi$-meson production at forward rapidity
in $p$$+$$p$ collisions at $\sqrt{s}$=510~GeV and  its energy dependence 
from $\sqrt{s}$=200 GeV to 7 TeV}

\newcommand{\abilene}{Abilene Christian University, Abilene, Texas 79699, USA}
\newcommand{\augie}{Department of Physics, Augustana University, Sioux Falls, South Dakota 57197, USA}
\newcommand{\banaras}{Department of Physics, Banaras Hindu University, Varanasi 221005, India}
\newcommand{\barc}{Bhabha Atomic Research Centre, Bombay 400 085, India}
\newcommand{\baruch}{Baruch College, City University of New York, New York, New York, 10010 USA}
\newcommand{\bnlcoll}{Collider-Accelerator Department, Brookhaven National Laboratory, Upton, New York 11973-5000, USA}
\newcommand{\bnlphys}{Physics Department, Brookhaven National Laboratory, Upton, New York 11973-5000, USA}
\newcommand{\caucr}{University of California-Riverside, Riverside, California 92521, USA}
\newcommand{\charlesczech}{Charles University, Ovocn\'{y} trh 5, Praha 1, 116 36, Prague, Czech Republic}
\newcommand{\chonbuk}{Chonbuk National University, Jeonju, 561-756, Korea}
\newcommand{\cns}{Center for Nuclear Study, Graduate School of Science, University of Tokyo, 7-3-1 Hongo, Bunkyo, Tokyo 113-0033, Japan}
\newcommand{\colorado}{University of Colorado, Boulder, Colorado 80309, USA}
\newcommand{\columbia}{Columbia University, New York, New York 10027 and Nevis Laboratories, Irvington, New York 10533, USA}
\newcommand{\czechtech}{Czech Technical University, Zikova 4, 166 36 Prague 6, Czech Republic}
\newcommand{\debrecen}{Debrecen University, H-4010 Debrecen, Egyetem t{\'e}r 1, Hungary}
\newcommand{\elte}{ELTE, E{\"o}tv{\"o}s Lor{\'a}nd University, H-1117 Budapest, P{\'a}zm{\'a}ny P.~s.~1/A, Hungary}
\newcommand{\eszterhazy}{Eszterh\'azy K\'aroly University, K\'aroly R\'obert Campus, H-3200 Gy\"ongy\"os, M\'atrai \'ut 36, Hungary}
\newcommand{\ewha}{Ewha Womans University, Seoul 120-750, Korea}
\newcommand{\fsu}{Florida State University, Tallahassee, Florida 32306, USA}
\newcommand{\gsu}{Georgia State University, Atlanta, Georgia 30303, USA}
\newcommand{\hanyang}{Hanyang University, Seoul 133-792, Korea}
\newcommand{\hiroshima}{Hiroshima University, Kagamiyama, Higashi-Hiroshima 739-8526, Japan}
\newcommand{\howard}{Department of Physics and Astronomy, Howard University, Washington, DC 20059, USA}
\newcommand{\ihepprot}{IHEP Protvino, State Research Center of Russian Federation, Institute for High Energy Physics, Protvino, 142281, Russia}
\newcommand{\illuiuc}{University of Illinois at Urbana-Champaign, Urbana, Illinois 61801, USA}
\newcommand{\inrras}{Institute for Nuclear Research of the Russian Academy of Sciences, prospekt 60-letiya Oktyabrya 7a, Moscow 117312, Russia}
\newcommand{\instpasczech}{Institute of Physics, Academy of Sciences of the Czech Republic, Na Slovance 2, 182 21 Prague 8, Czech Republic}
\newcommand{\isu}{Iowa State University, Ames, Iowa 50011, USA}
\newcommand{\jaea}{Advanced Science Research Center, Japan Atomic Energy Agency, 2-4 Shirakata Shirane, Tokai-mura, Naka-gun, Ibaraki-ken 319-1195, Japan}
\newcommand{\jyvaskyla}{Helsinki Institute of Physics and University of Jyv{\"a}skyl{\"a}, P.O.Box 35, FI-40014 Jyv{\"a}skyl{\"a}, Finland}
\newcommand{\kek}{KEK, High Energy Accelerator Research Organization, Tsukuba, Ibaraki 305-0801, Japan}
\newcommand{\korea}{Korea University, Seoul, 136-701, Korea}
\newcommand{\kurchatov}{National Research Center ``Kurchatov Institute", Moscow, 123098 Russia}
\newcommand{\kyoto}{Kyoto University, Kyoto 606-8502, Japan}
\newcommand{\lahorelums}{Physics Department, Lahore University of Management Sciences, Lahore 54792, Pakistan}
\newcommand{\lawllnl}{Lawrence Livermore National Laboratory, Livermore, California 94550, USA}
\newcommand{\losalamos}{Los Alamos National Laboratory, Los Alamos, New Mexico 87545, USA}
\newcommand{\lund}{Department of Physics, Lund University, Box 118, SE-221 00 Lund, Sweden}
\newcommand{\lyon}{IPNL, CNRS/IN2P3, Univ Lyon, Université Lyon 1, F-69622, Villeurbanne, France}
\newcommand{\maryland}{University of Maryland, College Park, Maryland 20742, USA}
\newcommand{\mass}{Department of Physics, University of Massachusetts, Amherst, Massachusetts 01003-9337, USA}
\newcommand{\michigan}{Department of Physics, University of Michigan, Ann Arbor, Michigan 48109-1040, USA}
\newcommand{\muhlenberg}{Muhlenberg College, Allentown, Pennsylvania 18104-5586, USA}
\newcommand{\myongji}{Myongji University, Yongin, Kyonggido 449-728, Korea}
\newcommand{\nara}{Nara Women's University, Kita-uoya Nishi-machi Nara 630-8506, Japan}
\newcommand{\natmephi}{National Research Nuclear University, MEPhI, Moscow Engineering Physics Institute, Moscow, 115409, Russia}
\newcommand{\newmex}{University of New Mexico, Albuquerque, New Mexico 87131, USA}
\newcommand{\nmsu}{New Mexico State University, Las Cruces, New Mexico 88003, USA}
\newcommand{\ohio}{Department of Physics and Astronomy, Ohio University, Athens, Ohio 45701, USA}
\newcommand{\ornl}{Oak Ridge National Laboratory, Oak Ridge, Tennessee 37831, USA}
\newcommand{\orsay}{IPN-Orsay, Univ.~Paris-Sud, CNRS/IN2P3, Universit\'e Paris-Saclay, BP1, F-91406, Orsay, France}
\newcommand{\peking}{Peking University, Beijing 100871, People's Republic of China}
\newcommand{\pnpi}{PNPI, Petersburg Nuclear Physics Institute, Gatchina, Leningrad region, 188300, Russia}
\newcommand{\riken}{RIKEN Nishina Center for Accelerator-Based Science, Wako, Saitama 351-0198, Japan}
\newcommand{\rikjrbrc}{RIKEN BNL Research Center, Brookhaven National Laboratory, Upton, New York 11973-5000, USA}
\newcommand{\rikkyo}{Physics Department, Rikkyo University, 3-34-1 Nishi-Ikebukuro, Toshima, Tokyo 171-8501, Japan}
\newcommand{\saispbstu}{Saint Petersburg State Polytechnic University, St.~Petersburg, 195251 Russia}
\newcommand{\seoulnat}{Department of Physics and Astronomy, Seoul National University, Seoul 151-742, Korea}
\newcommand{\stonybrkc}{Chemistry Department, Stony Brook University, SUNY, Stony Brook, New York 11794-3400, USA}
\newcommand{\stonycrkp}{Department of Physics and Astronomy, Stony Brook University, SUNY, Stony Brook, New York 11794-3800, USA}
\newcommand{\tenn}{University of Tennessee, Knoxville, Tennessee 37996, USA}
\newcommand{\titech}{Department of Physics, Tokyo Institute of Technology, Oh-okayama, Meguro, Tokyo 152-8551, Japan}
\newcommand{\tsukuba}{Tomonaga Center for the History of the Universe, University of Tsukuba, Tsukuba, Ibaraki 305, Japan}
\newcommand{\vandy}{Vanderbilt University, Nashville, Tennessee 37235, USA}
\newcommand{\weizmann}{Weizmann Institute, Rehovot 76100, Israel}
\newcommand{\wigner}{Institute for Particle and Nuclear Physics, Wigner Research Centre for Physics, Hungarian Academy of Sciences (Wigner RCP, RMKI) H-1525 Budapest 114, POBox 49, Budapest, Hungary}
\newcommand{\yonsei}{Yonsei University, IPAP, Seoul 120-749, Korea}
\newcommand{\zagreb}{Department of Physics, Faculty of Science, University of Zagreb, Bijeni\v{c}ka c.~32 HR-10002 Zagreb, Croatia}
\affiliation{\abilene}
\affiliation{\augie}
\affiliation{\banaras}
\affiliation{\barc}
\affiliation{\baruch}
\affiliation{\bnlcoll}
\affiliation{\bnlphys}
\affiliation{\caucr}
\affiliation{\charlesczech}
\affiliation{\chonbuk}
\affiliation{\cns}
\affiliation{\colorado}
\affiliation{\columbia}
\affiliation{\czechtech}
\affiliation{\debrecen}
\affiliation{\elte}
\affiliation{\eszterhazy}
\affiliation{\ewha}
\affiliation{\fsu}
\affiliation{\gsu}
\affiliation{\hanyang}
\affiliation{\hiroshima}
\affiliation{\howard}
\affiliation{\ihepprot}
\affiliation{\illuiuc}
\affiliation{\inrras}
\affiliation{\instpasczech}
\affiliation{\isu}
\affiliation{\jaea}
\affiliation{\jyvaskyla}
\affiliation{\kek}
\affiliation{\korea}
\affiliation{\kurchatov}
\affiliation{\kyoto}
\affiliation{\lahorelums}
\affiliation{\lawllnl}
\affiliation{\losalamos}
\affiliation{\lund}
\affiliation{\lyon}
\affiliation{\maryland}
\affiliation{\mass}
\affiliation{\michigan}
\affiliation{\muhlenberg}
\affiliation{\myongji}
\affiliation{\nara}
\affiliation{\natmephi}
\affiliation{\newmex}
\affiliation{\nmsu}
\affiliation{\ohio}
\affiliation{\ornl}
\affiliation{\orsay}
\affiliation{\peking}
\affiliation{\pnpi}
\affiliation{\riken}
\affiliation{\rikjrbrc}
\affiliation{\rikkyo}
\affiliation{\saispbstu}
\affiliation{\seoulnat}
\affiliation{\stonybrkc}
\affiliation{\stonycrkp}
\affiliation{\tenn}
\affiliation{\titech}
\affiliation{\tsukuba}
\affiliation{\vandy}
\affiliation{\weizmann}
\affiliation{\wigner}
\affiliation{\yonsei}
\affiliation{\zagreb}
\author{A.~Adare} \affiliation{\colorado} 
\author{C.~Aidala} \affiliation{\michigan} 
\author{N.N.~Ajitanand} \altaffiliation{Deceased} \affiliation{\stonybrkc} 
\author{Y.~Akiba} \email[PHENIX Spokesperson: ]{akiba@rcf.rhic.bnl.gov} \affiliation{\riken} \affiliation{\rikjrbrc} 
\author{R.~Akimoto} \affiliation{\cns} 
\author{M.~Alfred} \affiliation{\howard} 
\author{N.~Apadula} \affiliation{\isu} \affiliation{\stonycrkp} 
\author{Y.~Aramaki} \affiliation{\riken} 
\author{H.~Asano} \affiliation{\kyoto} \affiliation{\riken} 
\author{E.T.~Atomssa} \affiliation{\stonycrkp} 
% ...
\collaboration{PHENIX Collaboration}  \noaffiliation

\date{\today}

\begin{abstract}


The PHENIX experiment at the Relativistic Heavy Ion Collider has 
measured the differential cross section of $\phi$(1020)-meson 
production at forward rapidity in $p$$+$$p$ collisions at 
$\sqrt{s}=$510 GeV via the dimuon decay channel. The partial 
cross section in the rapidity and $p_T$ ranges $1.2<|y|<2.2$ and 
$2<p_T<7$ GeV/$c$ is $\sigma_\phi=[2.28 \pm 0.09\,{\rm (stat)} 
\pm 0.14\,{\rm (syst)} \pm 0.27\, {\rm (norm)}] \times 
10^{-2}$~mb.  The energy dependence of $\sigma_\phi$ 
($1.2<|y|<2.2, \; 2<\pt<5$ GeV/$c$) is studied using the PHENIX 
measurements at $\sqrt{s}=$200 and 510 GeV and the 
Large-Hadron-Collider measurements at $\sqrt{s}=$2.76 and 7 TeV. 
The experimental results are compared to various event generator 
predictions ({\sc pythia6, pythia8, phojet, ampt, epos3,} and 
{\sc epos-lhc}).

\end{abstract}

\maketitle

\section{Introduction}

The $\phi$(1020)-vector-meson production in \pp collisions was 
intensively studied by various experiments at different colliding 
energies and in different rapidity 
ranges~\cite{Drijard:1981ab,Daum:1981tw,Akesson:1982jg,Bobbink:1982kw,AguilarBenitez:1991yy,Alexopoulos:1995ru,Afanasev:2000uu,Abelev:2008aa,Adare:2010fe,Aaij:2011uk,Aamodt:2011zza,ALICE:2011ad,Abelev:2012hy,Aaltonen:2013bda,Adare:2014mgt,Aad:2014rca,Adam:2015jca,Adam:2017zbf}. 
It is the lightest bound state of $s$ and 
$\bar{s}$ quarks and is considered a good probe to study strangeness 
production in \pp collisions. Production of \pphi mesons from an initial 
nonstrange colliding system, such as \pp collisions, is substantially 
suppressed in comparison to $\omega$ and $\rho$ vector mesons due to the 
Okubo-Zweig-Iizuka (OZI) 
rule~\cite{Okubo:1963fa,Zweig:1981pd,Iizuka:1966fk}. 
The \pphi-meson production at low transverse momentum is dominated by 
soft processes and is sensitive to the hadronization mechanism, while 
hard processes become dominant at higher transverse momentum. In \pp 
collisions, the production of strangeness is in general not well 
described by generators such as \pythia, which tend to underestimate the 
production of strange 
particles~\cite{Aaij:2011uk,Abelev:2012jp,Khachatryan:2011tm,Skands:2014pea}. 
The study of \pphi-meson production in \pp collisions is an important 
tool to study quantum chromodynamics (QCD), providing data to tune 
phenomenological QCD models in which an interplay is mandatory between 
perturbative QCD calculations, used in particular for hard parton 
production dominant at higher \pt, and phenomenological QCD models, 
needed to describe the nonperturbative hadronization into strange 
hadrons like the $\phi$ meson.

In addition, recently, a long-range near-side angular correlation was 
observed in \pp collisions at Large-Hadron-Collider (LHC) 
energies~\cite{Khachatryan:2010gv,Khachatryan:2015lva,Aad:2015gqa}, 
which led to the observation of collectivity in \pp 
collisions~\cite{Khachatryan:2016txc}. This observation generated 
various explanations~\cite{Li:2012hc}, including those based on the 
color-glass-condensate (CGC) model~\cite{Dumitru:2010iy}, and collective 
hydrodynamic flow~\cite{Bozek:2010pb} or color 
reconnection~\cite{Bierlich:2015rha,Martin:2016igp}. Being the heaviest 
easily accessible meson made of light quarks, \pphi-meson production 
provides the largest lever arm accessible to study effects that scale 
with mass, as should be the case for collective 
effects~\cite{Werner:2013tya}.

The study of \pphi-meson production in \pp collisions can be an 
important tool to gain insight into new phenomena, such as long-range 
angular correlations, that would have a direct impact in the field of 
relativistic heavy-ion collisions. The \pphi-meson production is an 
excellent observable to probe the strangeness enhancement in the 
quark-gluon plasma created in heavy-ion 
collisions~\cite{PhysRevLett.48.1066,PhysRevLett.54.1122,KOCH1986167}.

We report the \pphi-meson-production cross section measured in \pp 
collisions at \s~=~510 GeV. The analysis uses a data sample of 
144.6~pb$^{-1}$ of integrated luminosity obtained by the PHENIX 
experiment in 2013. The cross section is averaged over the rapidity 
($y$) interval $1.2<|y|<2.2$ and reported in several bins of transverse 
momentum (\pt) in the range $2<\pt <7$~GeV/$c$. The results are compared 
to several model predictions \cite{Sjostrand:2004ef, Skands:2014pea, 
Bopp:1998rc, Werner:2013tya,Pierog:2013ria,Lin:2004en} and to the 
measurements previously reported by the PHENIX experiment at \s~=~200 
GeV~\cite{Adare:2014mgt} and by the LHC experiments measuring the 
\pphi-meson-production cross section at forward rapidity at \s~=~2.76 and 
7~TeV~\cite{Aaij:2011uk,Aamodt:2011zza,ALICE:2011ad,Abelev:2012hy,Adam:2015jca}.
Measurements from experiments at the Relativistic Heavy Ion Collider 
(RHIC) and the LHC allow extracting the energy dependence of the 
\pphi-meson-production cross section in the rapidity range $1.2<y<2.2$, which 
provide information to further constrain model predictions.


\section{Experimental setup}

A complete description of the PHENIX detector can be found in 
Ref.~\cite{Adcox:2003zm}. The results presented here are obtained by 
measuring the \pphi meson via its \mumu decay channel using both PHENIX 
muon spectrometers covering forward and backward pseudorapidities, 
$1.2<|\eta|<2.2$, and the full azimuth. 

Each muon arm spectrometer comprises hadron absorbers, a muon tracker
(MuTr), which resides in a radial field magnet, and a muon identifier
(MuID). The absorbers are situated in front of the MuTr to provide
hadron (mostly pion and kaon) rejection and are built of 19~cm of copper,
60~cm of iron, and 36.2~cm of stainless steel. The MuTr comprises three
sets of cathode strip chambers in a radial magnetic field with an
integrated bending power of 0.8 T.m.  The final component is the MuID,
which has five alternating steel absorbers and Iarocci tubes to further
reduce the number of punch-through hadrons misidentified as muons.  Muon
candidates are identified by reconstructed tracks in the MuTr matched to
MuID tracks that penetrate through to the last MuID plane.

Another detector system relevant to this analysis is the beam-beam 
counter (BBC), comprising two arrays of 64~\v{C}erenkov counters, located 
on both sides of the interaction point and covering the pseudorapidity 
$3.1<|\eta|<3.9$. The BBC system is used to measure the \pp collision 
vertex position along the beam axis ($z_{\rm vtx}$) with 2 cm resolution 
and to provide the minimum bias (MB) trigger.


\section{Data analysis}
\label{sec:DataAnalysis}

The results presented here are based on the data sample collected by 
PHENIX during the 2013 \pp run at \s~=~510~GeV. The BBC counters 
provide the MB trigger, which requires at least one hit 
in each of the BBCs. Events, in coincidence with the MB trigger, 
containing a muon pair within the acceptance of the spectrometer are 
selected by the level-1 dimuon trigger requiring that at least two 
tracks penetrate through the MuID to its last layer. A total of 
$5.3\times10^8$ dimuon triggered events are recorded, which 
corresponds to a sampled integrated luminosity of 144.6~pb$^{-1}$.

\subsection{Raw yield extraction}
\label{sec:RawYieldExtraction}

A set of quality assurance cuts is applied to the data to select 
$p$$+$$p$ events and muon candidates as well as to improve the 
signal-to-background ratio. Good $p$$+$$p$ events are selected by 
requiring that the collision occurs in the fiducial interaction region 
$|z_{\rm vtx}|<30$~cm as measured by the BBC. No selection is made on 
the event's charged particle multiplicity. The MuTr tracks are matched 
to the MuID tracks at the first MuID layer in both position and angle. 
In addition, the track is required to have more than a minimum number of 
possible hits in the MuTr (12 out of the maximum 16) and MuID (6 out of 
the maximum 10), and cuts on the individual track $\chi^2$ values are 
applied. Furthermore, there is a minimum allowed single muon momentum 
along the beam axis, $p_z$, which is reconstructed and energy-loss 
corrected at the collision vertex, of 2.4\GeVc corresponding to the 
momentum cut effectively imposed by the absorbers. Finally, a cut on the 
$\chi^2$ of the fit to the common vertex of the two candidate tracks 
near the interaction point is made.

 \begin{figure}
 \includegraphics[width=1.0\linewidth]{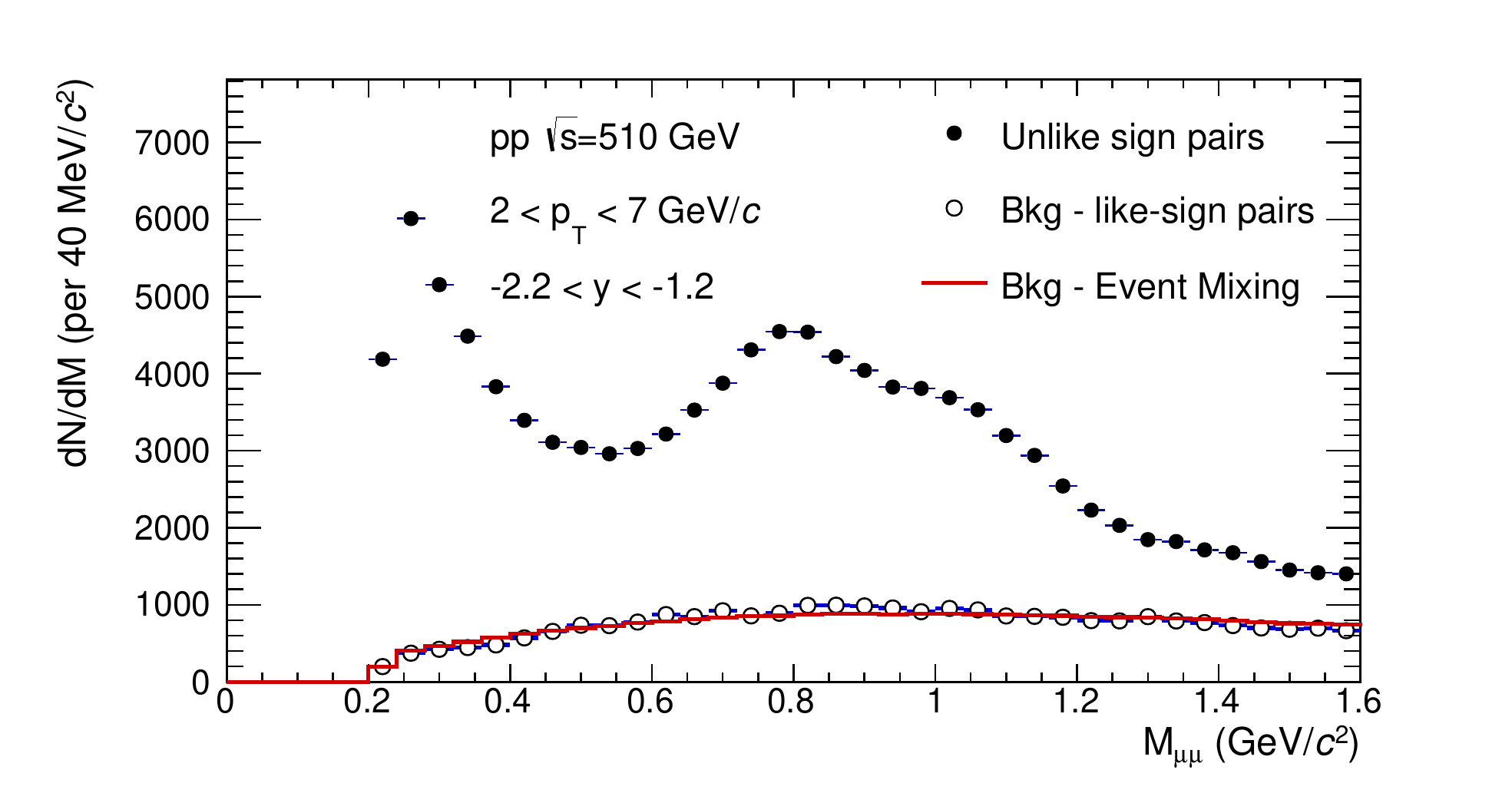}
 \caption{
Unlike-sign dimuon invariant mass spectrum before background subtraction 
(solid [black] circles) and uncorrelated background distribution estimated 
using like-sign pairs (open [black] circles) and the event-mixing technique 
(solid [red] curve ).
 \label{fig:MassPlot}}
 \end{figure}

The invariant mass distribution is formed by combining muon candidate 
tracks of opposite charges. This unlike-sign dimuon spectrum is composed 
of correlated and uncorrelated pairs. In the low-mass region (below 
$\approx$1.5\GeVcc) the correlated pairs arise from the two-body and 
Dalitz decays of the light neutral mesons $\eta$, $\rho$, $\omega$, 
$\eta'$ and $\phi$ as well as semi-muonic decays of correlated charmed 
hadrons (and beauty in a negligible contribution). The uncorrelated 
pairs are mainly coming from semi-muonic decays of pions and kaons and 
punch-through hadrons, and form the so-called combinatorial background. 
The ratio of \pphi-meson signal over combinatorial background is of the 
order of 0.7. This combinatorial background is estimated using two 
methods: the first one derives the combinatorial background from the 
distribution formed within the same event by the muon candidates of the 
same sign (like-sign pairs); and the second one derives the 
combinatorial background from the pairs formed by muon candidates of 
opposite charges (unlike-sign pairs) coming from different events 
(mixed-event). The normalization of the mass distribution of the 
combinatorial background using the same-event like-sign dimuon 
distributions ($N_{++}$ and $N_{--}$) is calculated as: $N_{\rm 
CB}=2\sqrt{N_{++}N_{--}}$.

The mixed-event like-sign dimuon mass distribution is normalized to the 
same-event like-sign combinatorial background distribution in the 
invariant mass range $0.2 - 2.5$\GeVcc. This factor is then used to 
normalize the mixed-event unlike-sign dimuon mass distribution. 
Figure~\ref{fig:MassPlot} shows the unlike-sign dimuon spectrum together 
with the combinatorial background estimated by both methods that agree 
within 15\% in the invariant mass range of interest 
(0.8~$<$~\massmumu~$<$~1.3\GeVcc).

\begin{figure}
 \includegraphics[width=1.0\linewidth]{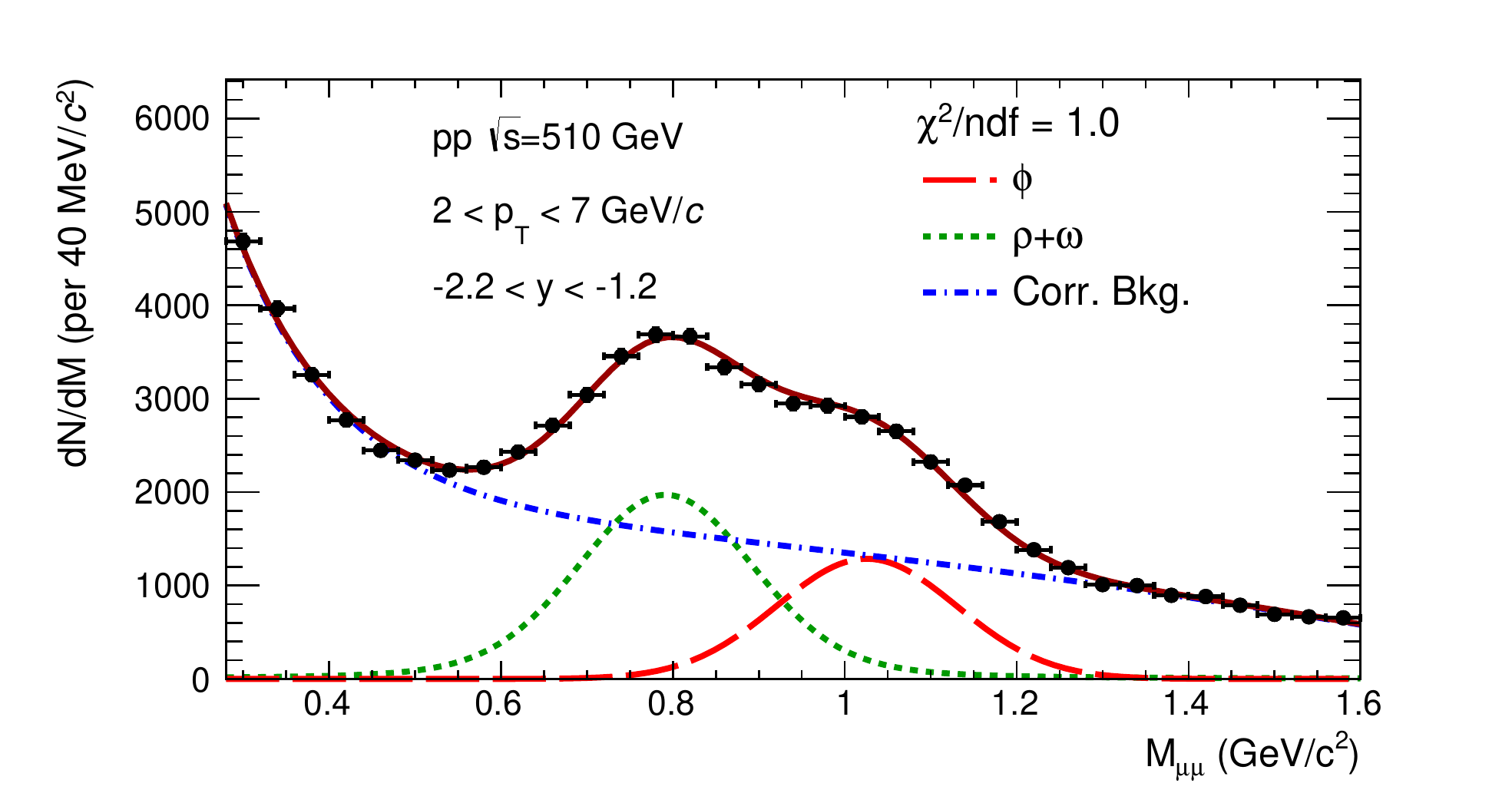}
\caption{Unlike-sign signal (solid [black] points) fitted by the sum of three 
components: \pphi meson (long dash [red] curve), $\rho+\omega$ mesons 
(short dash [green] curve), and correlated background (dot dash [blue] curve), 
see text for details.
\label{fig:MassPlotFitted}}
 \end{figure}

The signal invariant mass spectrum is extracted by first subtracting the 
uncorrelated combinatorial background spectra from the unlike-sign 
spectra. The signal spectra are then fitted to extract the \pphi 
contribution. The mass resolution of both muon spectrometers is 
estimated using Monte Carlo simulation to be 93 (94)~MeV/$c^2$ for the 
lowest \pt bin ($2<\pt<2.5 \GeVc$) and up to 114 (111)~MeV/$c^2$ for the 
highest \pt bin ($5<\pt<7 \GeVc$) for the negative (positive) 
pseudorapidity muon spectrometer. Those resolutions being greater than 
the natural widths of the \pphi and $\omega$, the two-body decay of 
\pphi and $\omega$ contributions are described by Gaussians while the 
two-body decay of the $\rho$-meson contribution is described by a 
Breit-Wigner distribution convoluted with a Gaussian. The contribution 
from $\rho$ dimuon decay is fixed by the assumption that the production 
cross section of $\rho$ and $\omega$ are related such as 
$\sigma_\rho=1.15\times \sigma_\omega$, as measured in 
Ref.~\cite{ALICE:2011ad} and used in previous PHENIX analysis related to 
\pphi-meson production in the dimuon decay channel~\cite{Adare:2014mgt, 
Adare:2015vvj, Adare:2015ema}. To evaluate the shape of the correlated 
background, a \pythia~\cite{Sjostrand:2006za} MB simulation followed by 
\geant~\cite{Brun:1994aa} transport and detector response simulation of 
the PHENIX detector is performed. The correlated background distribution 
is found to be well described by an exponential plus a polynomial of 
first order ($\chi^2$/ndf~$\le$~1). To summarize, eight free parameters 
are needed to describe the signal spectrum: two parameters for the \pphi 
and \rhomega signal normalizations, two parameters to describe relative 
changes of Gaussian widths and central masses with respect to simulation 
estimates and four parameters to describe the correlated background 
distribution and its normalization. The starting values of the free 
parameters describing the shapes of the different distributions are 
taken to be the ones from the Monte Carlo simulation.

Figure~\ref{fig:MassPlotFitted} shows the fit results for the entire \pt 
range at backward rapidity. Extracted peak positions and widths are 
found to be in good agreement with Monte Carlo simulations.
 
 \subsection{Detector acceptance and reconstruction efficiency}

\begin{figure}
  \includegraphics[width=1.0\linewidth]{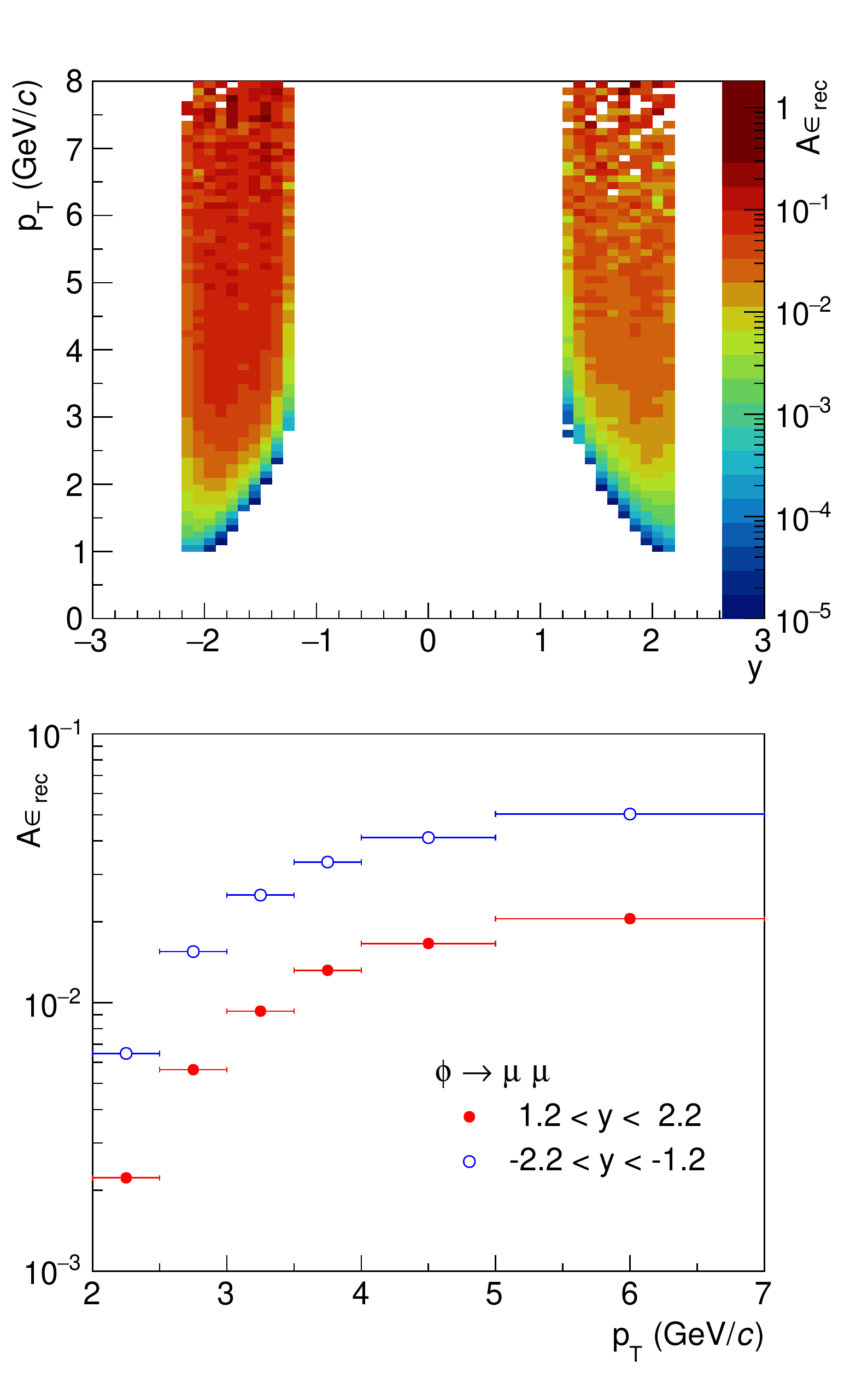} 
 \caption{\acc for \pphi detection in forward ($1.2< y< 2.2$) and 
backward ($-2.2< y< -1.2$) muon spectrometers (a) in the \pt-rapidity plane 
and (b) integrated in rapidity per spectrometer for each \pt bin 
considered in the analysis.
\label{fig:AccEffData}}
\end{figure}

The product of detector acceptance and reconstruction efficiency, \acc, 
of dimuon decays of \pphi mesons is determined by the full event 
reconstruction of the \pphi-meson signal run through a full \geant 
simulation of the 2013 PHENIX detector setup, and embedded in MB 
real-data. The \pt distribution of the simulated \pphi-meson signal is 
iteratively re-weighted to match the data \pt distribution, the initial 
\pt distribution being obtained from \pythiasix~\cite{Sjostrand:2006za} 
using tune ATLAS\_CSC~\cite{Sjostrand:2004ef}. The embedded simulated 
events are then reconstructed in the same manner as data with the same 
cuts applied as in the real data analysis. The \acc factor is extracted 
from the simulation as the ratio of reconstructed \pphi distribution 
over the generated one in the same kinematic range. 
Figure~\ref{fig:AccEffData} shows the \acc as a function of \pphi-meson 
\pt and rapidity. The main sources of the relative difference between 
both spectrometers \acc are different detection efficiencies of the MuTr 
and MuID systems and different amount of absorber material.

  \subsection{Differential cross section extraction}

The \pt-dependent differential cross section  is calculated according to:
\begin{equation}
\dsigdydpt = 
 \frac{N_{\rm raw}}{\acc\; \Delta\pt \; \Delta y \; {\rm BR_{\phi\rightarrow\mu^+\mu^-}} }\;\\
 \frac{\sigma_{\rm pp}^{\rm BBC}}{\epsilon_{\rm BBC}   \; N_{\rm MB}^{\rm BBC} },
\end{equation}
where ${\rm BR}_{\phi\rightarrow\mu^+\mu^-}=(2.87 \pm 0.19)\times 
10^{-4}$ is the branching ratio of \pphi decay to 
dimuon~\cite{Agashe:2014kda}. $N_{\rm raw}$ is the extracted \pphi raw 
yield for each \pt bin, $N_{\rm MB}^{\rm BBC}=4.16{\times}10^{12}$ is 
the number of sampled MB events. The BBC trigger samples a cross section 
of $\sigma_{pp}^{\rm BBC}=32.5{\pm}3.2$~mb in $p$$+$$p$ collisions, 
according to Vernier scans, however, it samples a larger fraction of the 
cross section when the collision includes a hard scattering 
process~\cite{Adare:2010xa}. Studies with high \pt $\pi^0$ yields show 
an increase of the luminosity scanned by the BBC by a factor of 
$1/\epsilon_{\rm BBC}$ , $\epsilon_{\rm 
BBC}=0.91{\pm}0.04$~\cite{Adare:2015ozj}. The inelastic cross sections 
given by \pythiaeight~\cite{Sjostrand:2014zea} for \s~=~500 and 510~GeV 
$p$$+$$p$ collisions differ by 0.3\%, therefore no correction or 
additional systematic uncertainty is added.
 
  \subsection{Systematic uncertainties}
 
The main source of systematic uncertainties in the signal extraction 
comes from the uncorrelated and correlated background distributions 
used.  To estimate this uncertainty, the extracted \pphi raw yields are 
compared using the following two methods; (1) the mixing and like-sign 
pair methods are separately used for subtraction of uncorrelated 
background and (2) the correlated background is fit by an exponential 
plus first-order polynomial and by an exponential plus second-order 
polynomial. The extracted \pphi raw yields are consistent among all 
different fit trials. The quadratic mean of the raw yields extracted 
from the trials is used as the central value, and the uncertainty on the 
central value is the quadratic mean of the uncertainties of all the 
trials. Table~\ref{tab:syst} summarizes the systematic uncertainties.

  \begin{table}[t]
  \caption{Systematic uncertainties associated with the differential 
cross section calculation.  \label{tab:syst}}
  \begin{ruledtabular}  \begin{tabular}{ccc} 
  Type & Origin & Value  \\ 
  \hline                                                              
A  &  Signal extraction  & 3--23\% \\
B &  \acc: \pt input distribution  &  2--8\% \\
B &  \acc: Rapidity input distribution  &  3--5\% \\
B &  \acc: Vertex width fluctuation  &  3.5\% \\
B &  \acc: MuID hit efficiency  &  4\% \\
B &  \acc: MuTr hit efficiency  &  2\% \\
B &  \acc: MuTr tracking efficiency  &  10\% \\
C &  MB trigger efficiency  &  10\% \\
C &  ${\rm Br}_{\phi\rightarrow\mu^+\mu^-}$  &  6.6\% \\
  \end{tabular} \end{ruledtabular}
\end{table}

Type-A is a point-to-point uncorrelated uncertainty which allows the 
data points to move independently with respect to one another and are 
added in quadrature with statistical uncertainties.  A systematic 
uncertainty equal to the difference between the central and the extreme 
values of the extracted yields accounts for the systematic uncertainty 
related to the background description as a whole.  The systematic 
uncertainty associated with the signal extraction method ranges from 3 
to 23\%, depending on the \pt bin and the muon spectrometer considered 
(negative/positive rapidity).

Type-B is a point-to-point correlated uncertainty which allows the data 
points to move coherently.  To evaluate the \acc systematic uncertainty, 
different \pt and rapidity input distributions of the simulated \pphi 
mesons are used. The \pt distribution is allowed to vary over the range 
of the data statistical uncertainty (statistical plus Type-A systematics 
uncertainties added in quadrature, see above), yielding an up to 8\% 
uncertainty. The rapidity distribution shapes given by five generator 
models (\pythiasix, \pythiaeight, \phojet, \epos and \eposlhc) are used 
as input rapidity distributions of the simulated \pphi mesons, resulting 
in up to 5\% uncertainty. The relative systematic uncertainty of 
acceptance caused by the fluctuation of vertex width is estimated to be 
3.5\% \cite{Oide:2012dla}. A 4\% uncertainty from the measured MuID tube 
efficiency and a 2\% uncertainty from MuTr chamber efficiency are 
assigned~\cite{Adare:2014mgt}. Simulation parameters are adjusted in 
order to reproduce the tracking efficiency observed in the data. While 
the relative tracking efficiency is validated using $J/\psi \rightarrow 
\mu \mu$ data, data-driven evaluation of the absolute tracking 
efficiency are not available. Therefore, we assign 10\% uncertainty for 
the absolute tracking efficiency as a conservative 
value~\cite{Oide:2012dla}.

Finally, Type-C is an overall normalization uncertainty, which allows 
the data points to move together by a common multiplicative factor. 
Type-C is composed of 10\% uncertainty assigned for the BBC cross 
section and efficiency uncertainties and a 6.6\% uncertainty from the 
measurement of ${\rm BR_{\phi\rightarrow\mu^+\mu^-}}$.

\section{Results}

The \pt-differential cross section is calculated independently for each 
muon arm, then the results are combined using the 
best-linear-unbiased-estimate method~\cite{Nisius:2014wua}. The \pt 
integrated ($2<\pt<7\GeVc$) cross section ${\rm d}\sigma_\phi/{\rm d}y$ 
is given in \tab{tab:InvariantCrossSectionYFinal}. Results obtained 
using the two muon spectrometers are consistent within uncertainties. 
Combining both arm results, the integrated cross section in the 
kinematic range $2<\pt<7\GeVc$ and $1.2<|y|<2.2$ is $\sigma_\phi=2.28 
\pm 0.09\, {\rm (stat)} \pm 0.14\,{\rm (syst)}\times 10^{-2}$~mb, to 
which a 12\% normalization uncertainty applies.

\begin{table}[htb]
\caption{ 
The \pphi-meson-production cross section ${\rm d}\sigma_\phi/ {\rm d}y$ in 
\pp collisions at \s=~510~GeV integrated in the transverse momentum 
range $2<\pt<7 \GeVc$. The first uncertainty represents the statistical 
and Type-A systematic uncertainties, while the second is the systematic 
uncertainty of Type-B and the third one is the additional $\pm$12\% 
Type-C normalization systematic uncertainty.
\label{tab:InvariantCrossSectionYFinal}
}
   \begin{ruledtabular} \begin{tabular}{cc}
  $y$ range &     ${\rm d}\sigma_\phi/ {\rm d}y$ (mb)  \\  
     \hline                                                           
$1.2<y<2.2$  &  $(2.13 \pm 0.14 \pm 0.16 \pm 0.26)\times 10^{-2}$   \\
$-2.2<y<-1.2$   &    $(2.46 \pm 0.12 \pm 0.18 \pm 0.30)\times 10^{-2}$   \\
$1.2<|y|<2.2$ &  $(2.28 \pm 0.09 \pm 0.14  \pm 0.27)\times 10^{-2}$  
  \end{tabular}   \end{ruledtabular}
\end{table}

\begin{table}[htb]
\caption{
The \pphi-meson-differential-production cross section 
${\rm d}^2\sigma_\phi/{\rm d}p_T {\rm d}y$ for $1.2<|y|<2.2$ 
in \pp collisions at \s=~510~GeV. {\it\~p}$_{T}$ is the \pt at 
which the data point is plotted (see text for details). The first 
uncertainty represents the statistical and Type-A systematic 
uncertainties, while the second is the systematic uncertainty of Type-B 
and the third one is the additional $\pm$12\% Type-C normalization 
systematic uncertainty.
\label{tab:InvariantCrossSectionFinal}
}
  \begin{ruledtabular}  \begin{tabular}{ccc}
  \pt range & \it{\~p}$_{T}$ &    $ {\rm d}^2\sigma_\phi/{\rm d}p_T {\rm d}y$   \\  [1ex]
 (\GeVc)  &    (\GeVc)& $\left[{\rm mb}/({\rm GeV/}c)\right]$ \\ 
    \hline
2.0--2.5 & 2.24  & $(2.16 \pm 0.17 \pm 0.23 \pm 0.26)\times 10^{-2}$\\
2.5--3.0 & 2.74  & $(1.20 \pm 0.05 \pm 0.12 \pm 0.14)\times 10^{-2}$\\
3.0--3.5 & 3.24  & $(6.26 \pm 0.36 \pm 0.61 \pm 0.75)\times 10^{-3}$\\
3.5--4.0 & 3.74  & $(2.70 \pm 0.20 \pm 0.30 \pm 0.32)\times 10^{-3}$\\
4.0--5.0 & 4.44  & $(1.06 \pm 0.07 \pm 0.11 \pm 0.13)\times 10^{-3}$\\
5.0--7.0 & 5.79  & $(1.97 \pm 0.19 \pm 0.20 \pm 0.24)\times 10^{-4}$\\
  \end{tabular}   \end{ruledtabular}
\end{table}

 \begin{figure}[htb] 
   \includegraphics[width=1.0\linewidth]{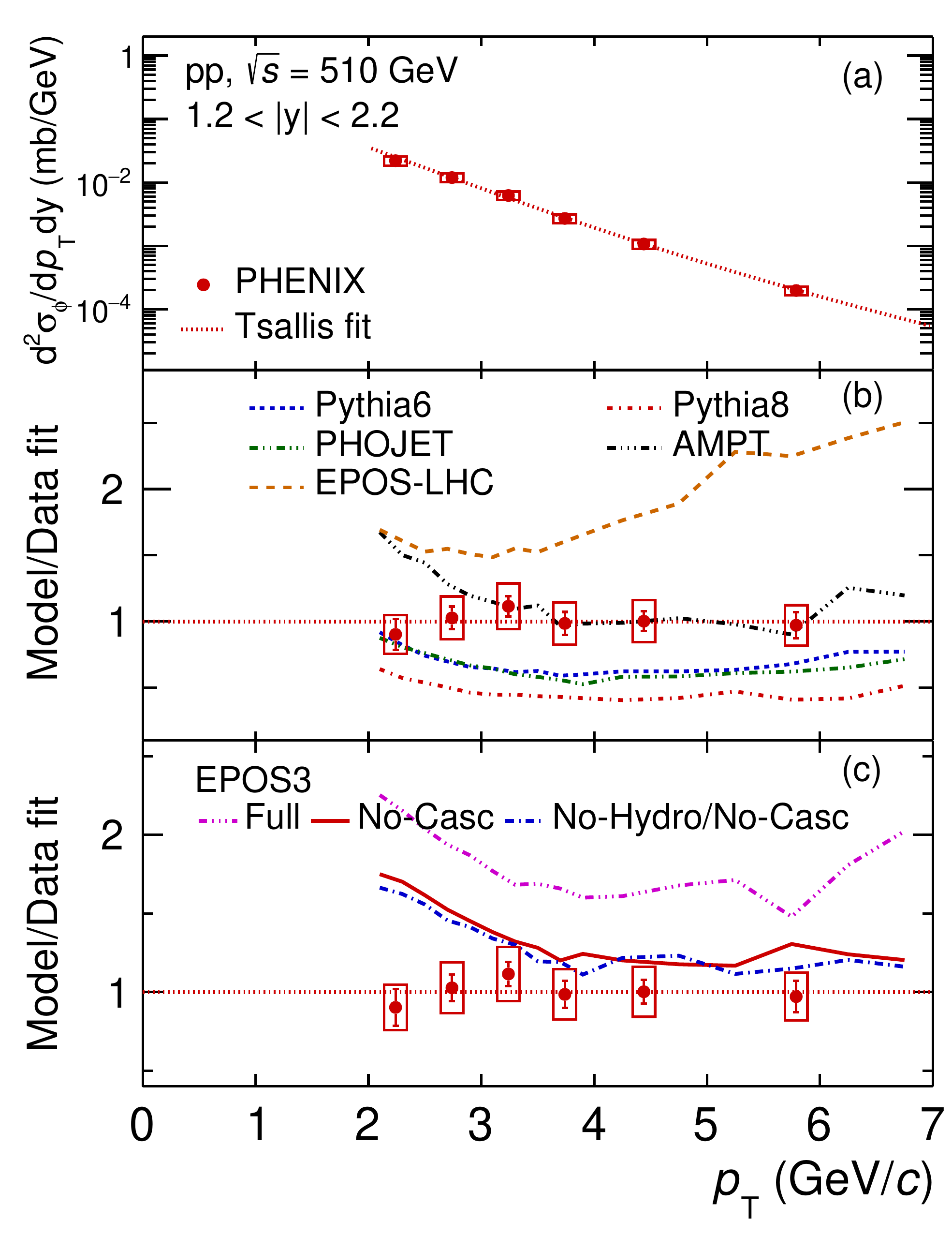} 
\caption{(a)  ${\rm d}^2\sigma_\phi / {\rm d}p_T {\rm d}y$ 
measurements in \pp collisions at \s~=~510~GeV fitted by a Tsallis 
function.  Error bars represent the statistical uncertainty and the boxes 
the Type-B and Type-C systematic uncertainties added in quadrature. 
(b) and (c)  Comparison between the data and predictions 
of six models (\pythiasix using tune ATLAS\_CSC, \pythiaeight using tune 
Monash2013, \phojet1.12, \eposlhc, \ampt~v1.26, and \epos.117) shown as 
the ratio of the model to the data fitted by a Tsallis function. 
(c) The data are compared to \epos predictions using three different 
options of the model (see text for details). 
\label{fig:d2sigdptdy_all_510}
}
\end{figure}

 \begin{figure*}[htb]
\begin{minipage}{0.42\linewidth}
  \includegraphics[width=0.99\linewidth]{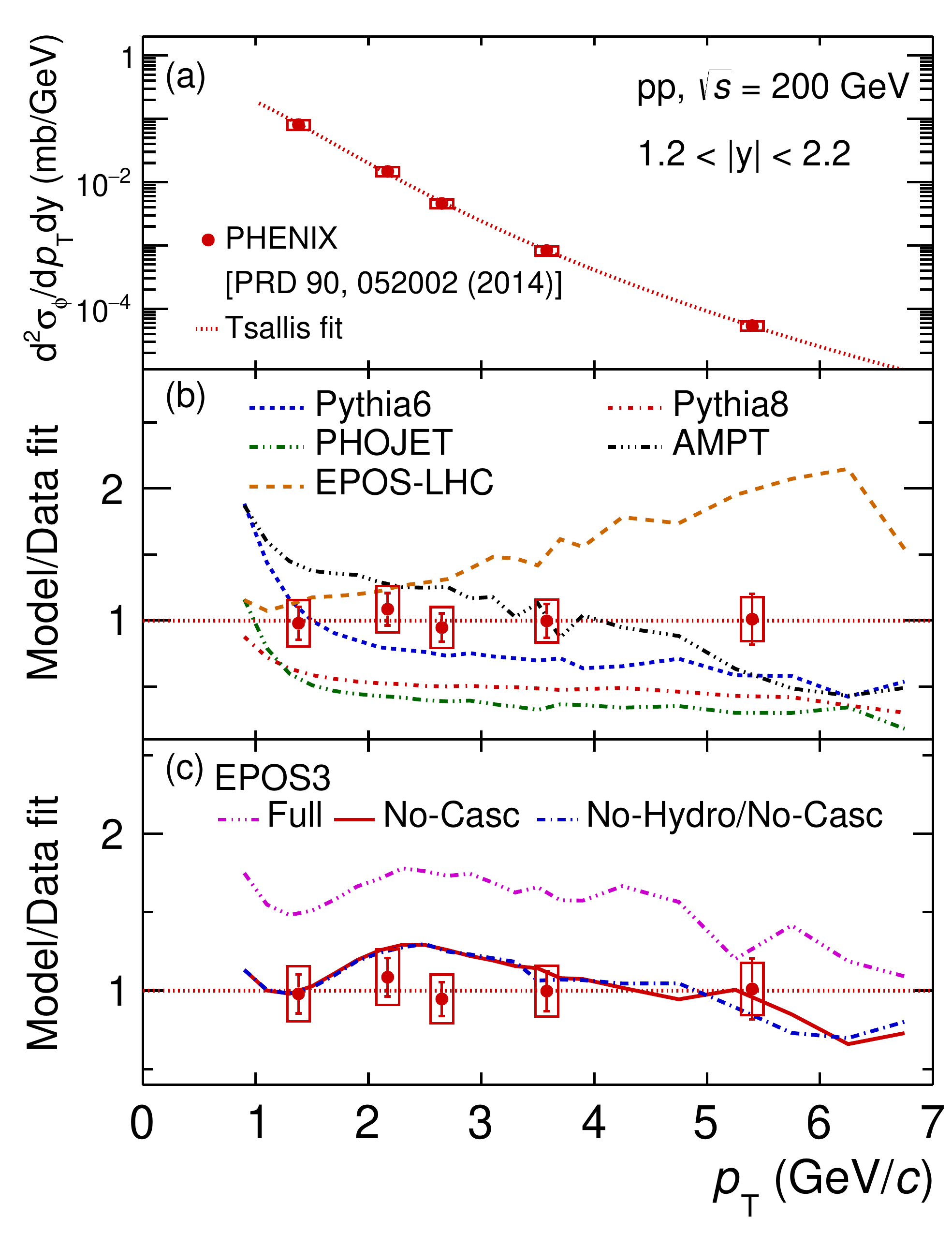} 
  \caption{Same as Fig.~\ref{fig:d2sigdptdy_all_510} for PHENIX 
measurement at \s~=~200~GeV~\protect\cite{Adare:2014mgt}.
   \label{fig:d2sigdptdy_phenix200}}
\end{minipage}
\hspace{0.3cm}
\begin{minipage}{0.42\linewidth}
   \includegraphics[width=0.99\linewidth]{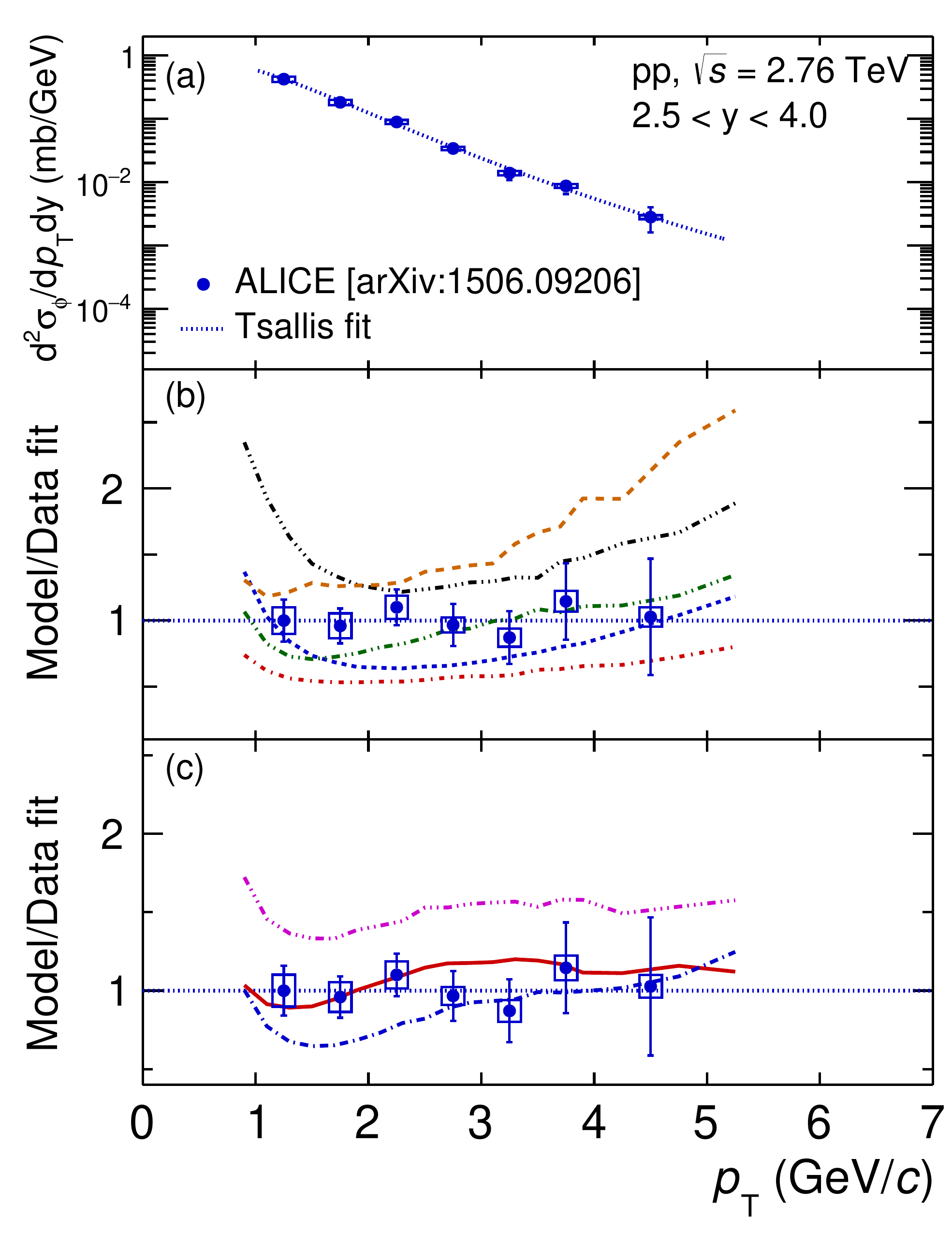} 
  \caption{Same as Fig.~\ref{fig:d2sigdptdy_all_510} for 
ALICE measurement at \s~=~2.76~TeV~\protect\cite{Adam:2015jca}. 
   \label{fig:d2sigdptdy_alice2760}}
\end{minipage}


\begin{minipage}{0.42\linewidth}
  \includegraphics[width=0.99\linewidth]{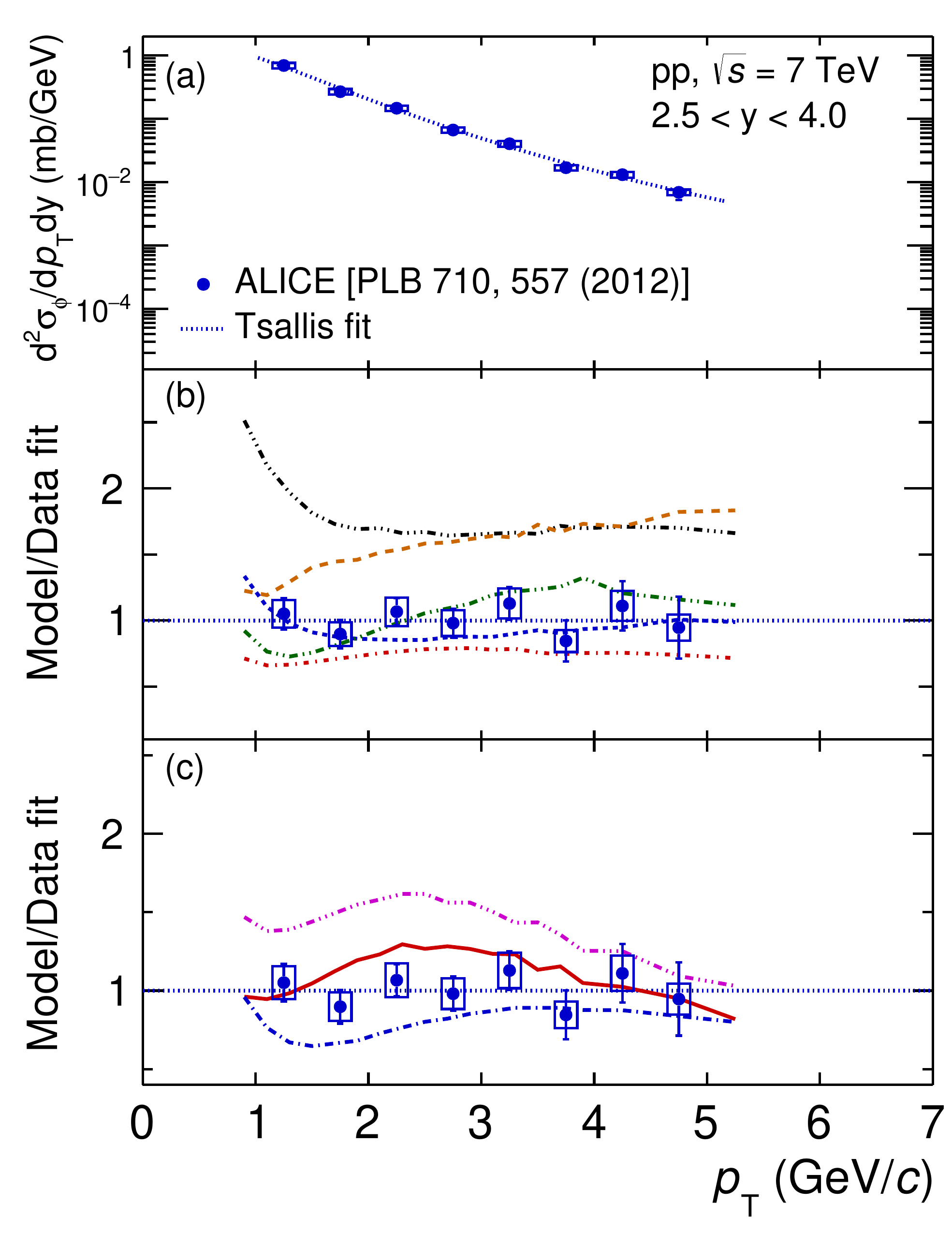} 
  \caption{Same as Fig.~\ref{fig:d2sigdptdy_all_510} for 
ALICE measurement at \s~=~7~TeV~\protect\cite{ALICE:2011ad}.
   \label{fig:d2sigdptdy_alice7000}}
\end{minipage}
\hspace{0.3cm}
\begin{minipage}{0.42\linewidth}
   \includegraphics[width=0.99\linewidth]{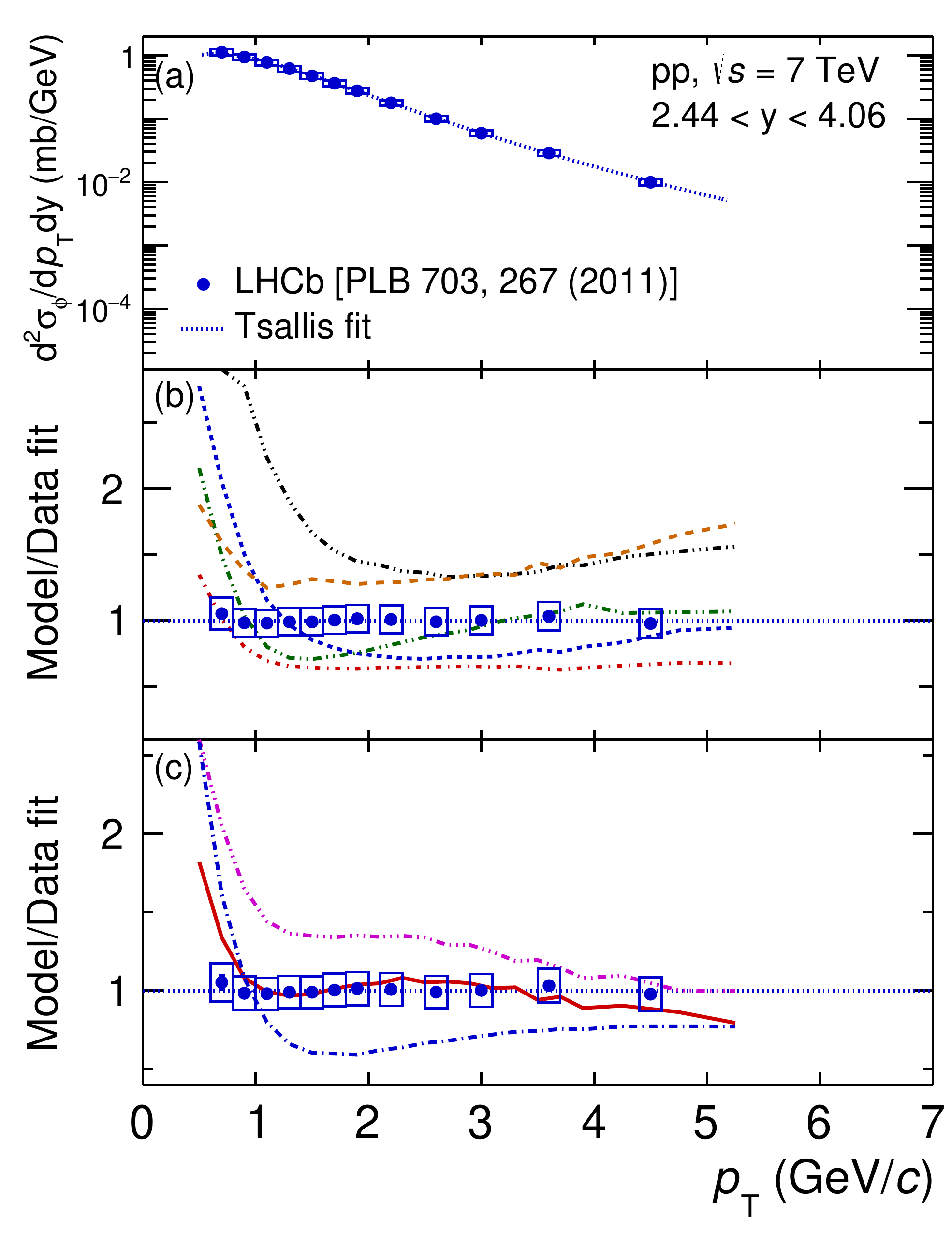} 
  \caption{Same as Fig.~\ref{fig:d2sigdptdy_all_510} for 
LHCb measurement at \s~=~7~TeV~\protect\cite{Aaij:2011uk}.
   \label{fig:d2sigdptdy_lhcb7000}}
\end{minipage}
\end{figure*}

The \pphi-meson-differential cross section as a function of \pt measured 
in \pp collisions at \s~=~510~GeV is shown in 
Fig.~\ref{fig:d2sigdptdy_all_510} and listed in 
Table~\ref{tab:InvariantCrossSectionFinal}. The data points are bin 
shifted in \pt using the Lafferty and Wyatt 
method~\cite{Lafferty:1994cj} to correct for the finite width of the \pt 
bins.

The data are fitted by a Tsallis function~\cite{Cleymans:2012ya} with a 
resulting $\chi^2$/ndf~=~0.66. The results are compared to calculations 
performed using six different generator models: 
\pythiasix~\cite{Sjostrand:2006za} using tune 
ATLAS\_CSC~\cite{Sjostrand:2004ef}, 
\pythiaeight.210~\cite{Sjostrand:2014zea} using tune 
Monash2013~\cite{Skands:2014pea}, \phojet1.12~\cite{Bopp:1998rc}, 
\epos.117~\cite{Werner:2013tya}, \eposlhc~\cite{Pierog:2013ria} and 
\ampt v1.26~\cite{Lin:2004en}. Data and models are compared as the ratio 
of the model prediction over the Tsallis fit of the data.

The \ampt simulation is done with the default \ampt model version 1.26 
(without string melting), where the initial conditions are determined by 
\hijing~\cite{Wang:1991hta}. Parton scattering is done using the Zhang's 
parton-cascade (ZPC) model~\cite{Zhang:1997ej}. The hadronization is 
accomplished using the Lund string fragmentation 
model~\cite{Andersson:1983jt,Andersson:1983ia}. The final state hadronic 
interactions are based on the ``a relativistic transport" (ART) 
model~\cite{Li:1995pra}. We used the set of parameters tabulated in 
Ref.~\cite{Xu:2011fe} describing both the charged particle 
distribution and elliptic flow measured in Au$+$Au collisions at RHIC. The 
Lund string fragmentation parameters are $a=0.5$ and $b=0.9$~GeV$^{-2}$, 
the QCD coupling constant is $\alpha_s=0.33$, and the screening mass is 
$\mu=3.2$~fm$^{-1}$, leading to a parton-scattering cross section of 
1.5~mb. Besides their production from the fragmentation of excited 
strings in the initial collisions, \pphi mesons can also be produced and 
absorbed from hadronic matter via various hadronic reactions 
(baryon-baryon, meson-baryon and meson-meson scatterings)~\cite{Lin:2004en}.

The \epos model includes, in addition to the description of the initial 
scattering based on a Gribov-Regge approach~\cite{Drescher:2000ha}, a 
viscous hydrodynamic expansion of the created system followed by a 
hadronization phase and a final state hadronic cascade using the \urqmd 
model~\cite{Bass:1998ca,Bleicher:1999xi}. In \epos, the hydrodynamic 
evolution and the hadronic cascade can be turned on or off, separately. 
The so-called ``Full" version of \epos includes hydrodynamic 
expansion of the created system followed by a final state hadronic 
cascade. The \epos ``No-Casc" version does not include the final state 
hadronic cascade and ``No-Hydro/No-Casc" has both hydrodynamic and the 
final state hadronic cascade turned off. The \eposlhc calculation 
presented in Fig.~\ref{fig:d2sigdptdy_all_510} is performed including a 
parameterized viscous hydrodynamic expansion of the created partonic 
system.

As shown in panels (b) and (c) of Fig.~\ref{fig:d2sigdptdy_all_510}, 
the experimental data are better reproduced by the \ampt model and by 
\epos without the hadronic cascade.  The \epos ``Full" and \eposlhc 
overestimate the \pphi-meson production, and \phojet and \pythia models 
tend to underestimate it by a factor of two.  A previous study of Monash2013 
tune of \pythiaeight showed that the calculated 
transverse-momentum spectra of \pphi mesons is overestimating the 
experimental data at very soft momenta (below $\sim$500 MeV/$c$) and 
underestimates it at higher momenta, the overall yield of \pphi mesons 
being correctly reproduced~\cite{Skands:2014pea}.

Additional calculations using the \ampt model with string melting 
(version 2.26) were performed. The \pphi-meson-production yield was found 
to be a factor of two higher than the one extracted using the default 
\ampt model with approximately the same \pt dependence.  For clarity, 
those calculations are not shown in Fig.~\ref{fig:d2sigdptdy_all_510}.


\section{Energy dependence of \pphi-meson production}

The PHENIX experiment previously measured the \pphi-meson cross section 
at forward rapidity and for $1<\pt<7\GeVc$ in \pp collisions at 
\s~=~200\GeV~\cite{Adare:2014mgt}. At the LHC, the ALICE experiment 
measured the \pphi-meson-production cross section via its dimuon decay channel 
in \pp collisions at forward rapidity $2.5<y<4.0$ and for 
$1<\pt<5\GeVc$ at \s~=~2.76~TeV~\cite{Adam:2015jca} and 
7~TeV~\cite{ALICE:2011ad}. Measurement of the \pphi-meson production was 
also performed via the $K^+K^-$ decay channel at midrapidity $|y|<0.5$ 
and for $0.4<\pt<6\GeVc$ at \s~=~7~TeV~\cite{Abelev:2012hy}. The LHCb 
experiment measured the inclusive \pphi-meson-production cross section in the 
$K^+K^-$ decay channel in the kinematic range $2.44<y<4.06$ and 
$0.6<\pt<5\GeVc$ in \pp collisions at \s~=~7~TeV~\cite{Aaij:2011uk}.

Figures~\ref{fig:d2sigdptdy_phenix200}--\ref{fig:d2sigdptdy_lhcb7000} 
show comparisons between 
${\rm d}^2\sigma_\phi /{\rm d}p_T {\rm d}y$ measurements at forward 
rapidities done by PHENIX at \s~=~200~GeV~\cite{Adare:2014mgt}, by ALICE 
at \s~=~2.76~TeV~\cite{Adam:2015jca} and 7~TeV~\cite{ALICE:2011ad} and 
by LHCb at \s~=~7~TeV~\cite{Aaij:2011uk}, respectively, along with model 
predictions. The \ampt model is in good agreement with the measured 
cross sections at both RHIC energies, but overestimates the production 
cross section at LHC energies, especially at 7~TeV. The \pythiasix and 
\phojet calculations at LHC energies are in better agreement with the 
data than at RHIC energies, where the models underestimate the measured 
production cross section. The \pythiaeight prediction underestimates the 
cross section for all four energies.

Panel (c) of Figs.~\ref{fig:d2sigdptdy_all_510}--\ref{fig:d2sigdptdy_lhcb7000} 
show the comparison between the 
measurements fitted by a Tsallis function and \epos using three 
different model settings (see above for details). The comparison of 
those results reveals the effect of the hydrodynamic expansion of the 
partonic system created in \pp collisions and of the final state 
hadronic cascade on the \pphi-meson production. The hydrodynamic 
evolution does not impact the \pphi-meson production at RHIC energies 
(``No-Casc" and ``No-Hydro/No-Casc" curves are almost identical on panel 
(c) of Figs~\ref{fig:d2sigdptdy_all_510}--\ref{fig:d2sigdptdy_lhcb7000}. 
A significant effect appears at 
\s~=~2.76~TeV and becomes stronger at 7~TeV where the \pphi-meson-production 
cross section increases by a factor of two for the \pt range 1--3 \GeVc 
when turning on the hydrodynamic evolution. The same behavior was 
already observed for the production of $\Lambda^0$, $K_s$ and $\Xi^\pm$ 
in \pp collisions at 7~TeV~\cite{Werner:2013tya}, showing that the flow 
effects increase with the mass of the particle. The final state hadronic 
cascade using the \urqmd model enhances the \pphi-meson-production cross 
section in the entire \pt range and for all collision energies. The 
\epos ``No-Casc" is the best configuration to reproduce the experimental 
data over the full collision energy range, while the addition of the 
\urqmd hadronic cascade overestimates the \pphi-meson production 
compared to the experimental data.


In the following, the \pphi-meson cross sections in the forward rapidity range 
$1.2<y<2.2$ at the different measured energies (0.2, 0.51, 2.76 and 
7~TeV) are presented. The \pt range is fixed to $2<\pt<5\GeVc$ which is 
the common range of all experimental measurements.

The cross sections measured by PHENIX in the kinematic range $1.2<y<2.2$ 
and $2<\pt<5\GeVc$ are:
\begin{itemize}
\item $\sigma_\phi(200\; {\rm GeV})=(1.10 \pm 0.17)\times 10^{-2}$~mb,
\item $\sigma_\phi(510 \;{\rm GeV})=(2.24 \pm 0.32)\times 10^{-2}$~mb,
\end{itemize}
where the uncertainties correspond to the quadratic sums of the 
statistical and systematic uncertainties.

\begin{figure}[htb] 
   \includegraphics[width=1.0\linewidth]{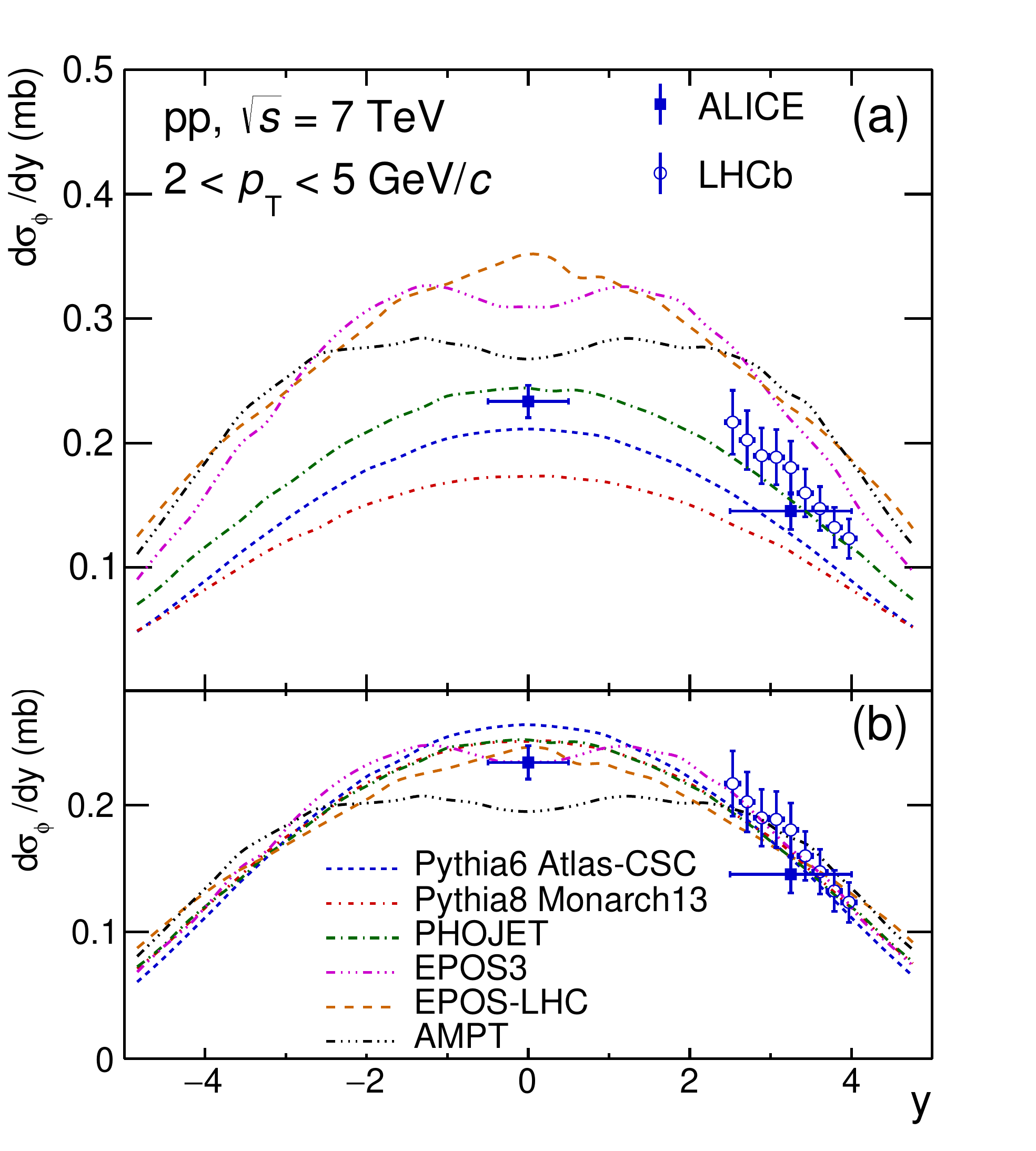} 
  \caption{ ${\rm d}\sigma_\phi/ {\rm d}y$ in \pp collisions at 7~TeV 
and for $2<\pt<5$\GeVc. (a) Comparison between LHC experiment 
measurements~\cite{ALICE:2011ad,Abelev:2012hy,Aaij:2011uk} and results 
of five simulations using \pythiasix, \pythiaeight, \phojet, \ampt, 
\epos and \eposlhc generator models. (b) Results of the fit of the 
measurements using the simulated ${\rm d}\sigma_\phi/ {\rm d}y$ shapes 
with normalization as the only free parameter.
  \label{fig:Phi_dsigdy_ALICE}
}
\end{figure}

The rapidity domains of the LHC measurements are different from those 
of PHENIX. Accordingly, to compare with PHENIX measurements the LHC 
measurements are extrapolated to the same rapidity coverage (i.e. $1.2<y<2.2$). The 
procedure followed here is to fit the LHC data points using the 
${\rm d}\sigma_\phi/{\rm d}y$ shapes obtained using the different models 
mentioned above, the only free parameter being the normalization of the 
simulated ${\rm d}\sigma_\phi/ {\rm d}y$ distributions. 
Figure~\ref{fig:Phi_dsigdy_ALICE} shows the LHC \pt integrated data 
points overlaid on the ${\rm d}\sigma_\phi/ {\rm d}y$ distributions 
obtained using \pythiasix, \pythiaeight, \phojet, \epos, \eposlhc and 
\ampt models at \s~=~7~TeV (a) before and (b) after the minimization 
procedure.

The LHC $ {\rm d}\sigma_\phi/ {\rm d}y$ at $1.2<y<2.2$ is calculated as 
the quadratic mean of the $ {\rm d}\sigma_\phi/ {\rm d}y$ from each of 
the model fits. The difference between the mean and the extreme value is 
taken as a systematic uncertainty, due to the rapidity shifting 
procedure, and added in quadrature to the experimental uncertainties. 
This uncertainty is 22.1\% for the 2.76~TeV measurement and 15.5\% at 
7~TeV. The obtained cross sections in $1.2<y<2.2$ and $2<\pt<5\GeVc$ 
at LHC energies are:
\begin{itemize}
\item $\sigma_\phi(2.76 \;{\rm TeV})=(1.15 \pm 0.28)\times 10^{-1}$~mb,
\item $\sigma_\phi(7 \;{\rm TeV})=(2.23 \pm 0.35)\times 10^{-1}$~mb.
\end{itemize}

\begin{figure}[htb] 
  \includegraphics[width=1.0\linewidth]{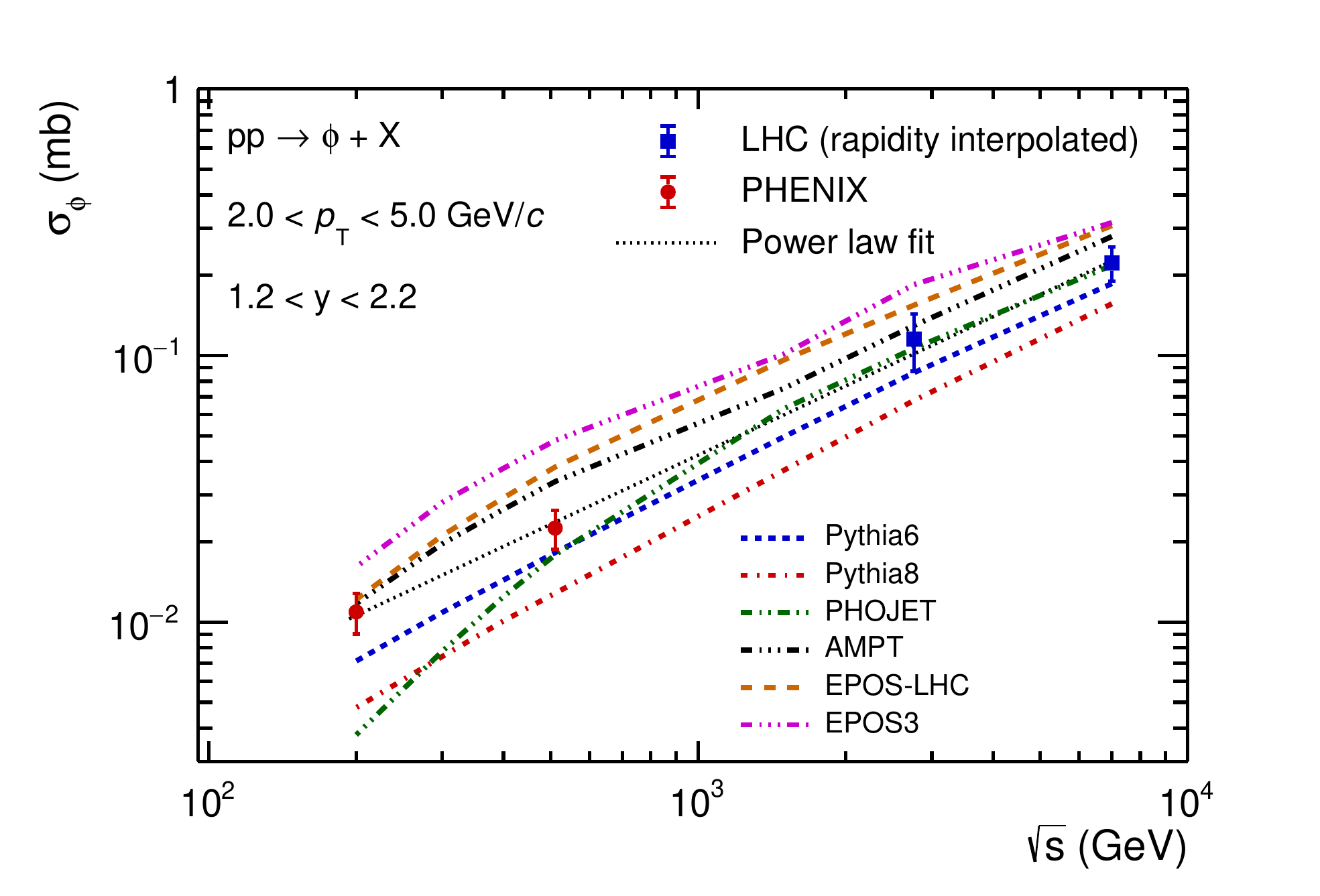} 
  \caption{ Partial-\pphi-meson-production cross section in $1.2<y<2.2$ and 
$2<\pt<5$\GeVc in \pp collisions versus the center-of-mass energy \s 
compared to different model predictions. The LHC data points are 
interpolated at the PHENIX forward rapidity, see text for details.
  \label{fig:Phi_versus_NRJ_all}
}
\end{figure}

Figure~\ref{fig:Phi_versus_NRJ_all} shows the energy dependence of the 
partial-\pphi-meson-production cross section integrated in $1.2<y<2.2$ and 
$2<\pt<5$\GeVc in \pp collisions compared to \pythiasix, \pythiaeight, 
\phojet, \ampt, \epos and \eposlhc model predictions.

 \begin{figure}[htb] 
  \includegraphics[width=1.0\linewidth]{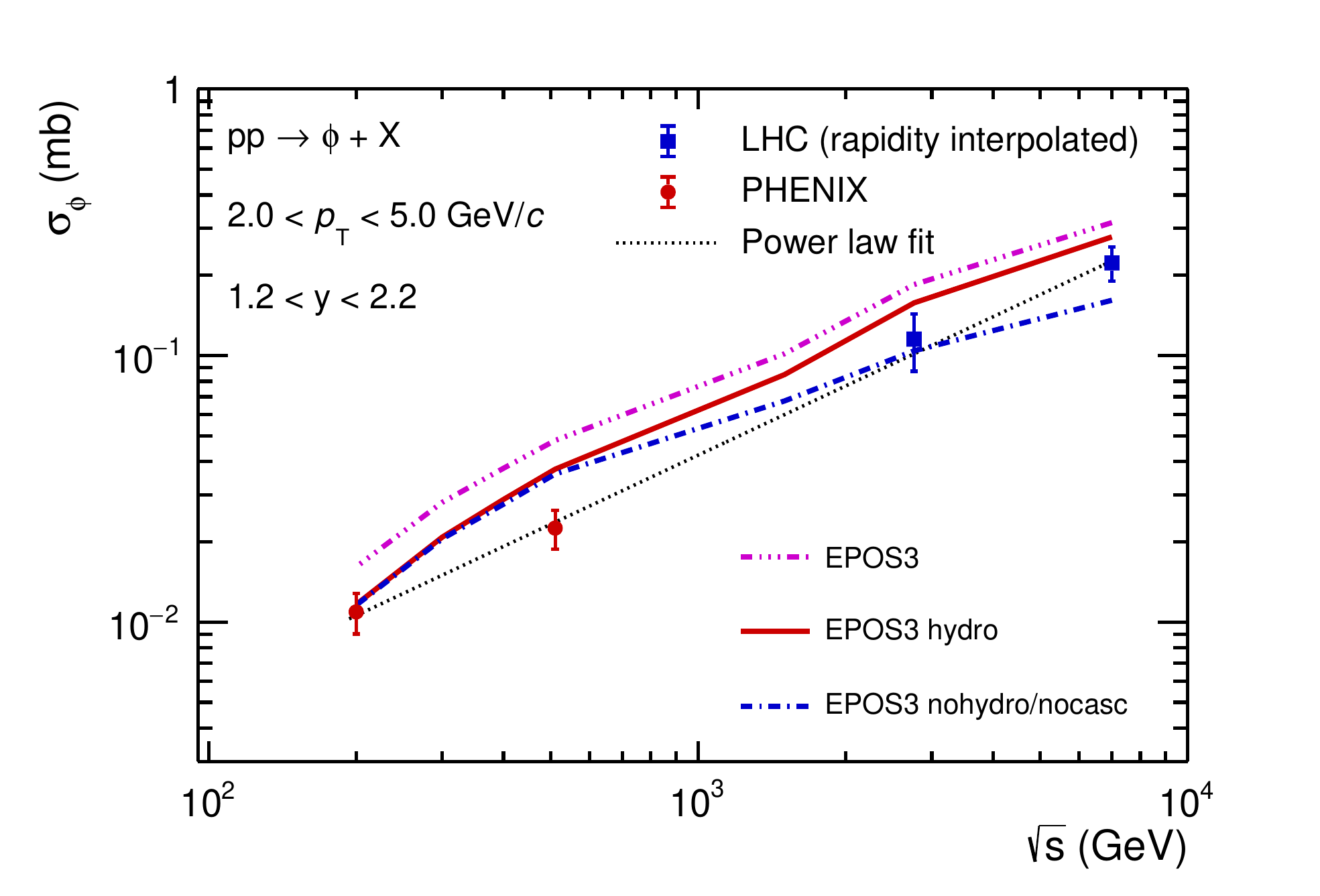} 
  \caption{Partial-\pphi-meson-production cross section in $1.2<y<2.2$ and 
$2<\pt<5$\GeVc in \pp collisions versus the center-of-mass energy \s 
compared to \epos model using different options. The LHC data points are 
interpolated at the PHENIX forward rapidity, see text for details.
  \label{fig:Phi_versus_NRJ_EPOS}
}
\end{figure}

The experimental measurements follow a power-law versus the colliding 
energy defined as $\sigma_\phi(s) \propto s^n,$ with $n=0.43{\pm}0.03$ 
(black dotted line in Fig.~\ref{fig:Phi_versus_NRJ_all}). The 
$\chi^2$/ndf of the power-law fit is 0.19.

The \phojet generator reproduces the partial \pphi-meson cross section 
correctly for LHC energies, but completely fails at RHIC energies. On 
the other hand, the \ampt model performs well at lower energies but 
overshoots the experimental data at 7~TeV. \pythiasix shows an energy 
dependence following a power law with exponent $n=0.43$, comparable to 
that of the data, but underestimates the cross section by $\sim$30\%. 
Accounting for hydrodynamic evolution of the partonic system makes \epos 
qualitatively and quantitatively better consistent with the data from 
both RHIC and LHC. The increasing effect of the hydrodynamic evolution of 
the system on the \pphi-meson production as the energy increases can 
clearly be seen in Fig.~\ref{fig:Phi_versus_NRJ_EPOS}. Also, the 
\pphi-meson enhancement caused by the hadronic cascade is approximately 
constant over the whole energy range, $\approx$20-30\%.

\begin{figure}[htb] 
  \includegraphics[width=1.0\linewidth]{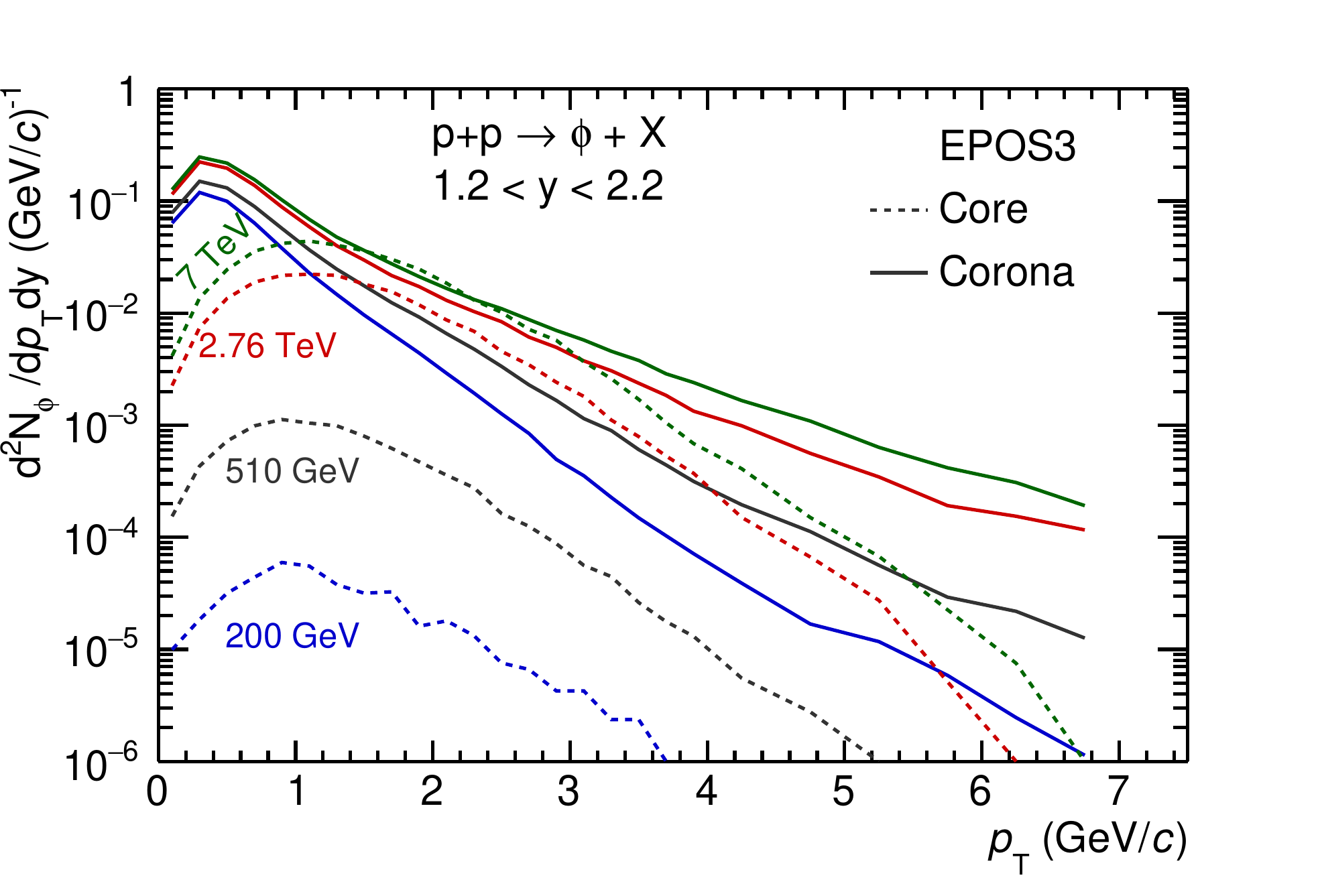} 
  \caption{Core (dashed curves) and Corona (solid curves) contributions to 
the production of \pphi mesons in \pp collisions at 0.2 [blue], 0.51 
[black], 2.76 [red] and 7~TeV [green].
  \label{fig:dNphidptdy_CoreCorona}
}
\end{figure}

In \epos, when the hydrodynamic evolution is turned off the hadrons are 
produced via string decays. On the other hand, when hydrodynamic 
calculation is included, the various string segments originating from 
the initial Pomerons are separated into two collections named ``core" 
and ``corona". The ``core" part will experience the hydrodynamic 
evolution while the segments in the ``corona" will leave the bulk matter 
and decay to hadrons. String segments are placed in the ``core" or 
``corona" depending on their transverse momenta and on the local string 
density~\cite{Werner:2013tya}. After its hydrodynamical evolution, the 
``core" hadronizes following the Cooper-Frye freeze-out procedure. 
Figure~\ref{fig:dNphidptdy_CoreCorona} shows the ``core" and the 
``corona" contributions to the production of \pphi mesons in \pp 
collisions for the four energies studied in this work. The contribution 
of the ``core" part increases with the colliding energy, being 
negligible compared to the ``corona" contribution at RHIC energies and 
of the same order of magnitude at LHC energies for $1<\pt<3$~\GeVc. 
The difference in the shape of the \pt distributions between the ``core" 
and the ``corona" part (shift from low to intermediate \pt) is due to 
the fact that in the ``core" the \pphi mesons are produced from ``fluid 
cells characterized by a radial flow velocities"~\cite{Werner:2013tya}. 
The heavier the particle is the more transverse momentum it receives 
from this mechanism.

\section{Summary and conclusions}

In summary, the \pphi-meson-production differential cross section is 
measured in \pp collisions at \s=510~GeV in the kinematic range 
$1.2<|y|<2.2$ and $2<\pt<7$\GeVc. The cross section integrated in \pt 
and averaged over positive and negative rapidities is 
$\sigma_\phi=[2.28 \pm 0.09\,{\rm (stat)} \pm 0.14\,{\rm (syst)} 
\pm 0.27\, {\rm (norm)}] \times 10^{-2}$~mb.
The measured \pt-differential cross section is compared to various model 
predictions based on \pythiasix, \pythiaeight, \phojet, \ampt, \epos and 
\eposlhc generators. The default \ampt model and the \epos model without 
hadronic cascade provide the best description of the data.

The energy dependence of the \pphi-meson-production cross section is 
studied in the kinematic range $1.2<y<2.2$ and $2<\pt<5$\GeVc, shifting 
LHC measurements to the same rapidity range as PHENIX measurements. The 
\epos model shows that the addition of the hydrodynamic evolution of the 
system induces an enhancement of the \pphi-meson production at the LHC 
energies for $1< \pt<3$~\GeVc whereas no effect is seen for RHIC 
energies. The LHC measurements tend to favor the scenario with 
hydrodynamic evolution of the system included in \epos showing a 
possible hint of collective effects in \pp collisions at high energy.

The \epos model shows that the hydrodynamic flow induces a shift from 
low to intermediate \pt of the produced \pphi mesons. A similar effect 
is obtained from tuning the color reconnection mechanism in 
\pythiaeight~\cite{Bierlich:2015rha, Martin:2016igp}. The study of the 
$\langle \pt \rangle$ as a function of the charged particle multiplicity 
produced in \pp collisions and its evolution versus the colliding energy 
would be a relevant observable of such effect, and would allow to 
discriminate between alternative models. In addition to the already 
published data at \s~=~2.76 and~7~TeV regarding the production of \pphi 
mesons at forward rapidity, the LHC experiments took data in \pp 
collisions at 5, 8, and recently 13~TeV where the effect should be even 
larger.

\section*{Acknowledgments}

We thank the staff of the Collider-Accelerator and Physics
Departments at Brookhaven National Laboratory and the staff of
the other PHENIX participating institutions for their vital contributions.  
We also thank K.~Werner and T.~Pierog for their valuable help in providing 
the \epos and \eposlhc software and for the fruitful discussions 
regarding the models.
We acknowledge support from the Office of Nuclear Physics in the
Office of Science of the Department of Energy,
the National Science Foundation, 
Abilene Christian University Research Council, 
Research Foundation of SUNY, and
Dean of the College of Arts and Sciences, Vanderbilt University 
(U.S.A),
Ministry of Education, Culture, Sports, Science, and Technology
and the Japan Society for the Promotion of Science (Japan),
Conselho Nacional de Desenvolvimento Cient\'{\i}fico e
Tecnol{\'o}gico and Funda\c c{\~a}o de Amparo {\`a} Pesquisa do
Estado de S{\~a}o Paulo (Brazil),
Natural Science Foundation of China (People's Republic of China),
Croatian Science Foundation and
Ministry of Science and Education (Croatia),
Ministry of Education, Youth and Sports (Czech Republic),
Centre National de la Recherche Scientifique, Commissariat
{\`a} l'{\'E}nergie Atomique, and Institut National de Physique
Nucl{\'e}aire et de Physique des Particules (France),
Bundesministerium f\"ur Bildung und Forschung, Deutscher
Akademischer Austausch Dienst, and Alexander von Humboldt Stiftung (Germany),
J. Bolyai Research Scholarship, EFOP, the New National Excellence
Program ({\'U}NKP), NKFIH, and OTKA (Hungary),
Department of Atomic Energy and Department of Science and Technology (India), 
Israel Science Foundation (Israel), 
Basic Science Research Program through NRF of the Ministry of Education (Korea),
Physics Department, Lahore University of Management Sciences (Pakistan),
Ministry of Education and Science, Russian Academy of Sciences,
Federal Agency of Atomic Energy (Russia),
VR and Wallenberg Foundation (Sweden), 
the U.S. Civilian Research and Development Foundation for the
Independent States of the Former Soviet Union, 
the Hungarian American Enterprise Scholarship Fund,
the US-Hungarian Fulbright Foundation,
and the US-Israel Binational Science Foundation.


%
 
\end{document}